 \documentclass[prc,twocolumn,floatfix,nofootinbib,showpacs,preprintnumbers]{revtex4}
  \usepackage{bm,amsmath,amssymb,graphicx} 
    \newcommand{\I}{\imath  }
    \newcommand{\D}{\textstyle{\rm d}}
    \newcommand{\E}{\textstyle{\rm e}}
\begin{document}

\title{Phenomenology of the  $pp\to pp\eta$  reaction close to threshold}

\author{A. Deloff}
 \email[]{deloff@fuw.edu.pl}
 \affiliation{Institute for Nuclear Studies, Ho\.{z}a 69, 00-681 Warsaw, Poland }

\date{\today}

  \begin{abstract}
  The recent high statistics measurement of the $pp\to pp\eta$ reaction at an
  excess energy Q=15.5~MeV has been analysed by means of partial wave decomposition
  of the cross section.
  Guided by the dominance of the final state  $\mbox{}^1S_0$  pp interaction
  (FSI), we keep only terms involving
  the FSI enhancement factor. The measured pp and $\eta$p effective mass
  spectra can be well reproduced by lifting the standard on-shell approximation
  in the enhancement factor and by allowing for
  a linear energy dependence in the leading 
  $\mbox{}^3 P_0\to\mbox{}^1 S_0,s$ partial wave amplitude.
  Higher partial waves seem to play only a marginal role.
  \end{abstract}


  \maketitle

 \section{\label{sec:1}Introduction} 

 In recent years major advances have been made in the experimental
 \cite{review,meyer,exp,calen,zupran,moskal}
 and theoretical \cite{eta_theo,vadimm,nakayama,etap_theo,kaiser}
 investigation of the near threshold meson production reactions in nucleon-nucleon collisions
 (for a comprehensive review cf. \cite{review}).
 With the advent  of medium-energy accelerators (ICUF, CELSIUS and COSY) 
 and the corresponding influx
 of high precision data on the total and differential cross section as well as
 the polarization observables, it has become possible to study the interaction of
 the flavor-neutral mesons (eg. $\pi^0,\,\eta$ and $\eta'$) with nucleons.
 A direct insight into such interaction can not be gained from meson-nucleon scattering experiments
 as the latter are impractical owing to the very short life time of these mesons.
 Naturally, the largest data base has been accumulated for the pions 
 \cite{meyer} but the bulk of data
 on $\eta$ production in proton-proton collisions has also expanded significantly
 \cite{exp, calen, zupran, moskal}.
 The $\eta$ meson, which is the next lightest non-strange member of the pseudoscalar octet,
 has focused considerable attention of the nuclear community since it was established that
 the $\eta$-N interaction was quite strong and attractive which  might lead to
 a possible existence of the $\eta$-nuclear bound states. 
 \par
  In the recent measurements of the $pp\to pp\eta$ reaction a very accurate determination
  of the four-momenta of both outgoing protons allowed for the full reconstruction
  of the kinematics of the final $\eta pp$ state. In consequence, these measurements
  provided in addition to the $\eta$ and the proton angular distributions, also the
  $pp$ and $\eta p$ effective mass distributions \cite{calen,zupran,moskal}. 
  The common feature of the near-threshold
  meson production in proton-proton collisions
  is the dominance of the very strong proton-proton final state interaction.
  The Monte Carlo simulations as well as direct calculations reveal 
  that for small excess energies the inclusion of FSI 
  enhances the cross section by more than an order of magnitude. This effect is also
  clearly visible in the effective mass distributions: as a prominent peak close 
  to threshold in the pp effective mass distribution or as a bump near the end-point
  in the $\eta p$ effective mass distribution. 
  The description in terms of a simple model in which
  a constant $\eta$ production amplitude is multiplied by an on-shell proton-proton
  FSI enhancement factor is only qualitatively correct,
  but at a  quantitative level turns out to be insufficient to explain the 
  experimental pp and $\eta$p effective mass distributions. 
  The contribution from higher partial waves and the final state $\eta$p 
  interaction have been indicated as the possible sources of this discrepancy. 
  In comparison with pp interaction, however, the role of $\eta$-N interaction 
  is much less important and taking as a rough measure the ratio of the $\eta$p and
  pp scattering lengths squared we may expect an effect at the level of about 1\%.
  A full three-body calculation based on the hyperspherical harmonics method
  accounting for all pair-wise final state interactions 
  corroborates the above estimate
  showing that the $\eta$p interaction modifies significantly only the total
  $\eta$ production cross section as function of the excitation energy whereas the distortion 
  of the shapes of
  the effective mass distributions does not exceed the current experimental uncertainties.
  The hyperspherical harmonics method will be presented elsewhere \cite{HH} and here
  instead we would like to confine our attention to a purely phenomenological
  description of the effective mass spectra at the lowest available excitation
  energy equal Q=15.5~MeV. At such a low Q only a few partial waves in both initial
  and final states are expected to participate in the $\eta$ creation process which
  substantially facilitates the interpretation of the experimental findings.
  Among the possible amplitudes the 
  partial wave amplitude in which the two protons are deposited in
  the $\mbox{}^1S_0$  state plays a dominant role because 
  it is proportional to the large FSI enhancement factor. Therefore, 
  in the first approximation, in the cross section
  it should be sufficient to retain  the square of the modulus
  of the dominant amplitude plus the interference terms that would contain the large FSI enhancement
  factor. However, after integrating over the angles the interference terms vanish and
  give no contribution to the pp effective mass distribution. This implies that 
  there is really no room for improvements left unless a weak energy dependence is admitted in
  the dominant partial wave amplitude. We demonstrate in this paper that with 
  this rather modest assumption both pp and $\eta$p 
  effective mass distributions can be quite well reproduced.   
  \par
  The plan of our presentation is as follows. In the next Section the FSI enhancement
  factor is revisited. We argue that owing to a too steep a fall of the on-shell 
  enhancement factor the approximate form thereof should be abandoned if favor of
  the full off-shell expression. Having established the best form of the enhancement 
  factor we discuss the partial wave expansion
  of the $pp\to \eta pp$ transition amplitude in a quest of an approximate 
  expression for the cross section that would be valid for the lowest excitation energies.
  Finally, in Section III we verify our simple model by 
  presenting the comparison with experiment.

 \section{\label{sec:2}Theoretical framework}       
 \subsection{Derivation of the enhancement factor}
 Since the proton-proton final-state interaction is believed to be the 
 dominant ingredient in the description of 
  the $pp\to pp\eta$ reaction close to threshold, 
 it is logical that we first wish to reexamine  the FSI problem 
 to make sure that adequate measures  have been taken
 to obtain the ultimate solution.
 The basic idea how to account for final state interaction was 
 put forward 50 years ago by Fermi \cite{Fermi}, Watson \cite{Watson}, Migdal \cite{Migdal}
 and others (for a review cf.  \cite{Gold,Omnes,Giles})  and is based on the
 observation that in many processes the interaction responsible for carrying the system from 
 the initial to the final state is of such a  short range that
 in the first approximation may be regarded as point like.
 As a prototype one may consider a meson (x) production reaction $NN\to NNx$. To generate
 the meson mass m in nucleon-nucleon collision a large momentum transfer is required
 between the initial and the final nucleons, which is typically of the order
 $\sqrt{Mm}$, with $M$ being the nucleon mass. The corresponding ''range''
 of the production interaction is therefore much shorter than the range of the
 interaction between the two final state nucleons. 
 Although it is perfectly true that 
 the final state NN interaction significantly  distorts the
 NN wave function but in the transition 
 matrix element the contribution from all but the smallest NN separations will be 
 strongly suppressed and the main effect may be attributed to
 the change of  the normalization of the wave function at zero separation. 
 If the non-interacting NN pair is described by a plane wave
 $\E^{-\I\,\bm{k}\cdot\bm{r}}$, where $\bm{k}$ is the relative NN momentum
 ($\hbar=c=1$ units are used hereafter), to account for
 final state interaction the latter must be replaced in the transition matrix element by 
 the complete NN wave function $\Psi^-(\bm{k},\bm{r})^\dagger$ 
 satisfying outgoing spherical wave
 boundary condition at infinity. 
 Nevertheless, for a {\it point-like interaction}, we may set 
 \begin{equation}
 \Psi^-(\bm{k},\bm{r})^\dagger \approx
 \E^{-\I\,\bm{k}\cdot\bm{r}} \; C(k)  
 \label{AA1}
 \end{equation}
 in the matrix element so that the final state interaction will be accounted
 for by multiplying the transition matrix element by the enhancement factor,
 defined as
\begin{equation}
\label{AA2}
 C(k) \equiv  \displaystyle\lim_{r\to 0} \;
 \Psi^+(-\bm{k},\bm{r}) /\E^{-\I\,\bm{k}\cdot\bm{r} }. 
 \end{equation}
The factor $|C(k)|^2$ that appears in the cross section 
represents the ratio of  two probabilities: one of finding the 
interacting NN pair at zero separation,  while the other probability is associated with 
non-interacting particles. By construction, when the final state interaction is turned off,
the enhancement factor will be equal to unity. Expanding both, the numerator and the denominator
on the right hand side of \eqref{AA2} in partial waves, we have
\begin{equation}
 C(k) =  \displaystyle\lim_{r\to 0}\dfrac
 { \sum_{\ell=0}^\infty\; (2\ell+1)\,\I^{-\ell}\,
 \psi_\ell(k,r)/r\;P_\ell(\hat{\bm{k}}\cdot\hat{\bm{r}}) }
 { \sum_{\ell=0}^\infty\; (2\ell+1)\, \I^{-\ell}\;
 j_\ell(kr)\,P_\ell(\hat{\bm{k}}\cdot\hat{\bm{r}}) },
\label{A2a}
\end{equation}
where $\psi_\ell(k,r)\sim r^{\ell+1}$ for small $r$, $j_\ell(kr)$ is spherical Bessel function and $P_\ell(\hat{\bm{k}}\cdot\hat{\bm{r}})$ denotes Legendre polynomial. 
Clearly, in the limit $r\to 0$ in \eqref{A2a}, all higher partial
waves will be suppressed by the centrifugal barrier, and
only the contribution the from s-wave survives. Thus, we obtain a simple formula
\begin{equation}
 C(k) =  \psi_0(k,0)'/k,
\label{C1}
\end{equation}
where prime denotes derivative with respect to $r$, and, 
as apparent from \eqref{C1}, the enhancement factor is 
determined by the slope of the wave function at the origin.
To find this slope we must know the NN s-wave interaction and
for simplicity we shall in the following assume that the 
latter takes the form of a spherically symmetric radial potential.
The shape of this potential may be arbitrary but it must be of a
short range. Given the NN potential,
we can integrate  outward the appropriate wave equation, containing
both the nuclear and the Coulomb potential, 
generating numerically a regular solution $u_0(k,r)$ (i.e. vanishing at the origin)
whose derivative satisfies the boundary condition
\begin{equation}
 u_0(k,0)'=C_0(\eta)\, k,
\label{C2}
\end{equation}
where $\eta$ denotes the Sommerfeld parameter and $C_0(\eta)^2=2\pi\eta/[\exp(2\pi\eta)-1]$ is
the Coulomb barrier penetration factor.
The sought for physical solution $\psi_0(k,r)$ occurring in \eqref{C1} 
which is also regular,
is necessarily proportional to $u_0(k,r)$, and, more explicitly, we have
\begin{equation}
\psi_0(k,r)=[C(k)/C_0(\eta)]\;u_0(k,r). 
\label{C3}
\end{equation}
Now, all we need to calculate  $C(k)$ is the asymptotic expression for
the physical wave function. For $r=R$ with R much bigger than the range of the 
nuclear potential, the physical wave function takes the form  
\begin{equation}
	\begin{split}
\psi_0(k,R)&= [C(k)/C_0(\eta)]\,u_0(k,R)=\\
           &= F_0(\eta,kR)+f_0(k)\,H^+_0(\eta,kR),\\
\label{C9}
        \end{split} 
\end{equation}
where $H^+_0(\eta,kR)= G_0(\eta,kR)+\I\, F_0(\eta,kR)$
with $G_0(\eta,kR)$ and  $F_0(\eta,kR)$ being the standard Coulomb
wave functions defined in \cite{Abram}, 
and $f_0(k)=\sin\delta\,\E^{\I\delta}$ denotes the s-wave scattering 
amplitude with $\delta$ being the s-wave phase shift.
 The differentiation of \eqref{C9} with respect to R, provides us with
 a second condition for the derivatives but
 it should be noted that  $u_0(k,R)$ and $u(k,R)'$ 
 occurring in these two matching conditions
are to be regarded as known quantities. Indeed,  they are fully specified by 
the boundary condition at the origin \eqref{C2} and can be 
either calculated analytically, or obtained by numerical methods.  
Therefore, what we end up with are two algebraic equations in which the two unknowns
are the enhancement factor
$C(k)$ and the scattering amplitude $f_0(k)$ and the respective solutions,   
can be conveniently written as
\begin{equation}
C(k)=\dfrac{k\,C_0(\eta)}{w[H^+_0(\eta,kR),u_0(k,R)]},
\label{C10}
\end{equation}
and
\begin{equation}
f_0(k)=-\dfrac{w[  F_0(\eta,kR),u_0(k,R)]}
              {w[H^+_0(\eta,kR),u_0(k,R)]},
 \label{C10a}
 \end{equation}
 where the symbol $w[f,g]$ denotes the Wronskian 
 defined as $w[f,g]\equiv fg'-f'g$. 
 We wish to recall that  
 the specific Wronskian involving the regular solution and the outgoing wave solution
 present in the denominator of  \eqref{C10}  and  \eqref{C10a}  
 has been referred to as the Jost function  \cite{Gold}.
 It is worth noting that unlike the scattering amplitude \eqref{C10a} which  
 can be expressed solely in terms of the logarithmic
 derivative of the regular solution, the enhancement factor \eqref{C10} depends separately upon 
 both, $u$ and $u'$, 
 and therefore the calculation of $C(k)$ requires off-shell information. In particular, 
 a phase shift equivalent transformation of the potential may change the enhancement factor
 by orders of magnitude.  
 Formula \eqref{C10} may be cast to the form
 \begin{equation}
 \label{t10}
 C(k)= C_{WM}(k)
  \; \frac{(-k^2)\,C_{0}(\eta)^2}
 {w[F_{0}(\eta,kR),u_0(k,R)]}, 
 \end{equation}
 where 
 \begin{equation}
 \label{t10a}
 C_{WM}(k)= \frac{\E^{\I\delta}\sin{\delta}}{k\,C_{0}(\eta)},
 \end{equation}
 is the familiar Watson-Migdal factor  \cite{Gold}, depending solely 
 upon the on-shell quantities, whereas 
 the Wronskian occurring in \eqref{t10} 
  represents the off-shell correction.%
 \par
 Formula \eqref{C10} deserves some further comments. 
 It is easy to check that when  both, the Coulomb and the
 strong interaction are switched off, the enhancement factor  \eqref{C10} 
 goes to unity.
 When the nuclear potential alone is set equal to zero, we have $u_0(k,R)=F_0(\eta,kR)$ and
 \eqref{C10} yields $C(k)=C_0(\eta)$.
 Clearly, these are the right limits but they 
 could not have been obtained  with the Watson-Migdal factor alone
 which implies the importance of the off-shell correction. 
 Nevertheless, close to threshold 
  the off-shell factor varies slowly with energy,
  and the Watson-Migdal factor usually makes a good approximation. %
  It should be kept in mind, however, that in result of
 multiplying the cross section by the Watson-Migdal factor the overall
 normalization is lost precluding an absolute calculation.
 \par
 Some authors choose to simplify further the Watson-Migdal enhancement factor \eqref{t10a} by
 applying a Coulomb modified effective range formula for the phase shift. 
 Thus, for the pp case, 
 the phase shift can be obtained e.g. from the expression  \cite{pp}
 \begin{equation}
	 \begin{split}
 C_0^2(\eta)\,k\,\cot{\delta} +2k\,\eta\, h(\eta)=&\\
  = -1/a+b\,k^2/2& -P\,k^4/(1+Qk^2),
 \label{t11}
 \end{split}
 \end{equation}
 with $h(\eta)={\text Re}\, \psi(1+\I\,\eta)-\log(\eta)$ where $\psi$ is the
 logarithmic derivative of the gamma function. In \eqref{t11} 
 $a=-7.83~fm$ and $b=2.8~fm$ denote, respectively, 
 the experimental pp scattering
 length and the effective range and the remaining two parameters 
 ($P=0.73~fm^{3}$ and $Q=3.35~fm^{2}$) are related to a specific
 NN potential  \cite{Nijm}. 
 The approximation  \eqref{t11} has been popular in ceratin quarters
 and used in conjunction with Watson-Migdal formula \eqref{t10a} 
 for not very large k, results in a good approximation to the enhancement
 factor owing to a rather fortuitous cancellation of
 errors associated with different approximations. 
 \subsection{Examples}     
 We are going now to illustrate the results obtained in the preceding
 subsection by explicit calculations carried out for three
 phenomenological NN potentials which are: the delta-shell $V_D$,
 a Gaussian $V_G$, and the soft-core Reid $V_R$ potential  \cite{Reid}.
 Among these three potentials only Reid potential has a repulsive
 short-range component while the two remaining ones are purely attractive. 
 The reason for selecting
 these particular shapes is that they exhibit different behavior
 in the close to the origin region: the delta-shell potential vanishes
 for small $r$,
 while a Gaussian potential shows maximum strength at $r=0$,
 and, finally,  the Reid potential is singular at the origin,
 i.e. $V_R \to \infty$.
 Our purpose is to examine how this very different off-shell
 behavior influences the enhancement factor properties. Thus,
 as our first example we shall consider the delta-shell potential specified
 by a range $R$ and a dimensionless depth parameter $s$, defined
 by the formula
 \begin{equation}
 M\, V_{D}(r) = -(s/R) \, \delta(r-R),
 \label{e1}
 \end{equation}
 where the values of the parameters are  $R=1.84~fm$ and $s=0.906$.
 Although this potential is not realistic, its main advantage lies
 in its simplicity: both the phase shift and the enhancement
 factor may be for this case obtained in an analytic form. 
 A simple calculation gives the phase shift
 \begin{equation}
 \label{e2}
 \tan\delta = \frac{(s/\rho)\,F_{0}(\eta,\rho)^{2}}
 {1-(s/\rho)\,F_{0}(\eta,\rho)\,G_{0}(\eta,\rho)}
 \end{equation}
 where $\rho=kR$, and, respectively, the enhancement factor
 \begin{equation}
 \label{e3}
 C(k) = \frac{C_{0}(\eta)}
 {1-(s/\rho)\,F_{0}(\eta,\rho)\,H^+_{0}(\eta,\rho)}.
 \end{equation}
 It is apparent from \eqref{e3} that when the nuclear potential is
 switched off by setting $s$ equal to zero the enhancement factor
 reduces to the Coulomb factor $C_{0}(\eta)$.%
 \par
 The Gaussian potential $V_{G}(r)=V_{0}\,\exp{(-r^{2}/R^{2}})$ has also
 two parameters, the depth $V_0$ and the range $R$,
 whose values are $V_{0}=-31~MeV$ and $R=1.8~fm$ and 
 in this case the the solution of the wave equation will be obtained
 numerically. Finally, we consider the fully realistic
 Reid soft core-potential  \cite{Reid}, given in the form of a superposition of
 three Yukawa potentials, which also requires numerical treatment.%
 \par
\begin{figure}
 \begin{center}
 \includegraphics[scale=0.4]{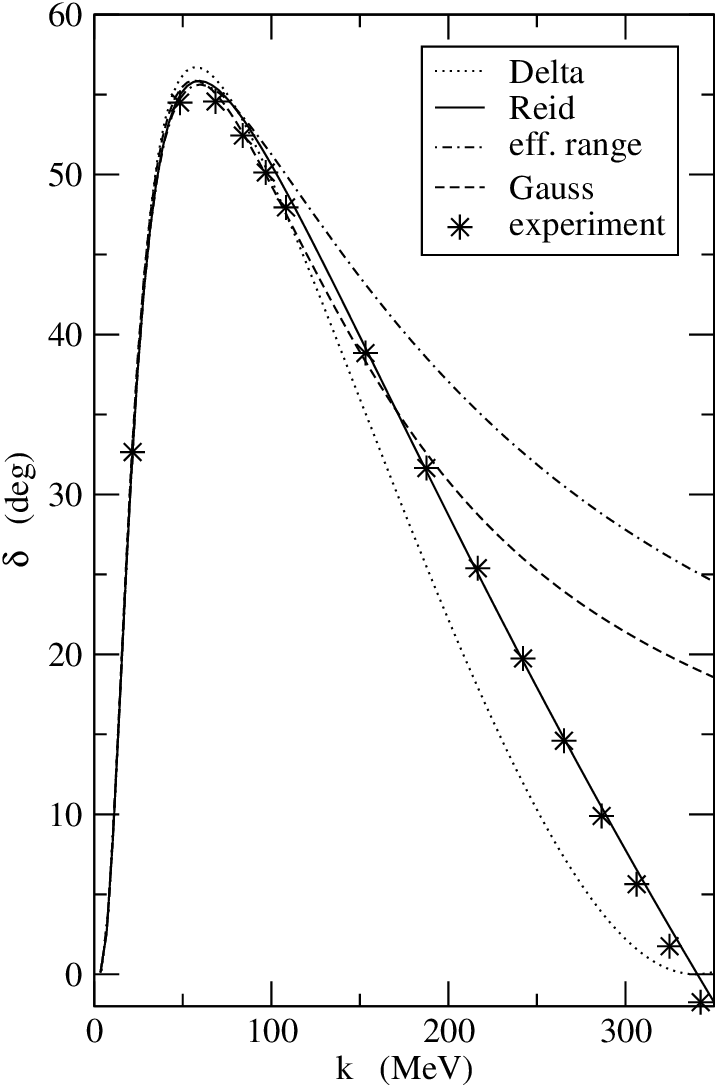}
 \caption{ Momentum dependence of $\mbox{}^1 S_0$ phase shift for different pp potentials.}
 \label{fig:1}
 \end{center}        
 \end{figure}
\begin{figure}
 \begin{center}
 \includegraphics[scale=0.4]{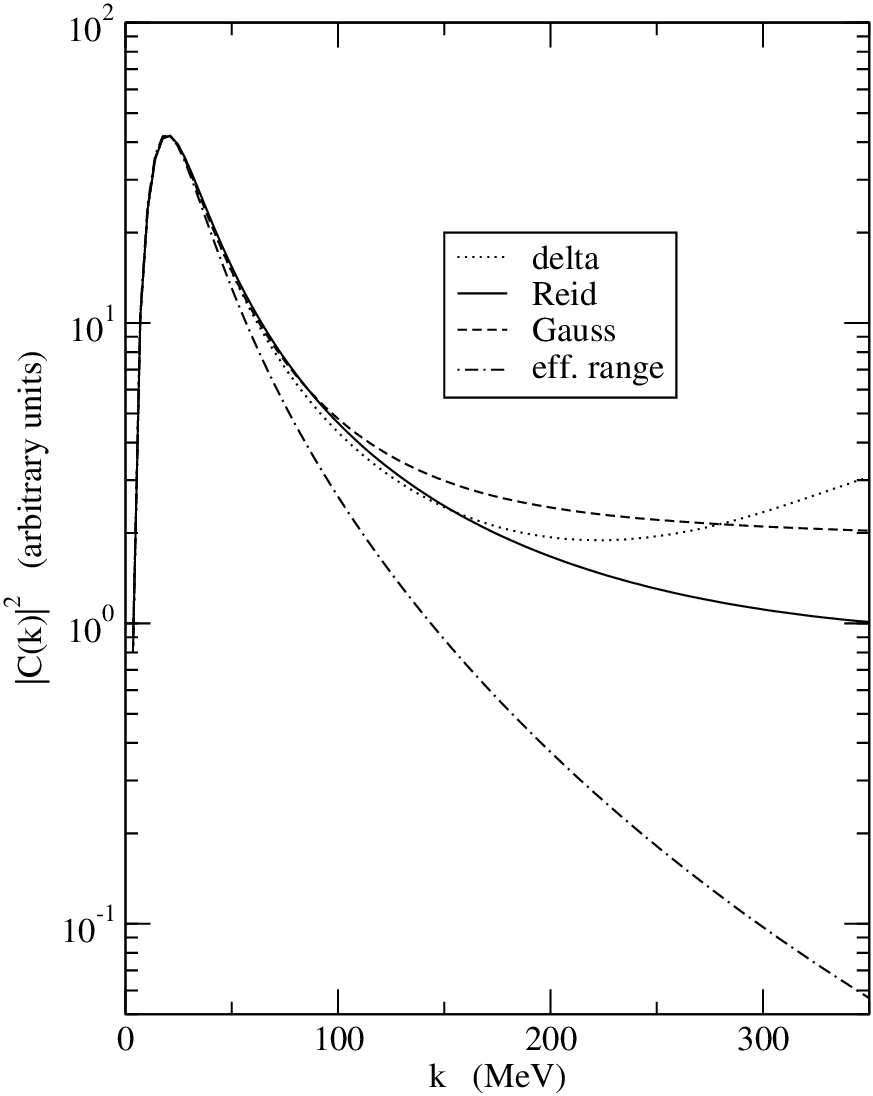}
 \caption{Enhancement factor $C(k)|^2$ vs. k for different pp potentials. The
 dot-dashed curve has been obtained form \eqref{t10a} and \eqref{t11}.}
 \label{fig:2}
 \end{center}        
 \end{figure}

 The results of our computations are presented in Figs.~\ref{fig:1} and \ref{fig:2}.
 In Fig.~\ref{fig:1} we show the calculated pp phase shift vs c.m. momentum as 
 obtained from the different potentials. They are compared with the 
 data and we can see that up to about 150~MeV all models agree well
 with experiment. For bigger momenta the situation is less
 satisfactory, except for the Reid potential whose performance is
 still very good. 
 \printfigures
 Clearly, among the considered potentials
 only the Reid potential has a repulsive core and
 therefore is capable of reproducing the crossover
 from positve to negative values in the phase shift at 340~MeV. 
 In Fig.~\ref{fig:2} we compare the enhancement factors calculated
 from  formula \eqref{t10} for different potentials.
 We stipulated the normalization as to get in all cases the same  
 height at the maximum. 
 It is evident from Fig.~\ref{fig:2} that very good
 agreement between different potential models results  sustains for momenta up to
 about 150~MeV indicating that, apart from normalization, different
 off-shell properties have little impact on the shape of 
 the enhancement factor. 
 This can be easily understood since close to threshold effective range approximation
 is usually adequate and all NN potentials are devised in such a way that they
 are capable of reproducing the experimentally determined effective range parameters. 
 The main consequence of conducting a
 phase shift equivalent transformation is that 
 the asymptotic wave function and its derivative acquire a common factor
 preserving thereby the shape of the enhancement factor.
 The differences appearing at larger momenta are not surprising
 as the different models do differ there also on-shell, as apparent from
 Fig.~\ref{fig:1}. In particular, as seen from Fig.~\ref{fig:2}, 
 the effective range formula makes a reliable approximation for k
 not bigger than about 80~MeV.
 With the current high precision data, however, this approximation is 
 insufficient causing  a  too rapid fall-off of the enhancement factor.
 Indeed, already for the excitation energy as low as Q=15.5~MeV 
 and the maximal momentum $k_{max}= 120~MeV$,   
 $|C_{WM}(k_{max})|^2$  calculated    from \eqref{t11} 
 is by  a factor of 2  smaller  from  $|C(k_{max})|^2$ obtained from
 \eqref{t10}, with both functions normalized to  yield the same height at the peak.   
 In consequence, with the on-shell approximation the relative momentum distribution
 in the $pp\to pp\eta$  reaction has a too rapid fall off.
 \subsection{The cross section of the reaction $pp\to pp\eta$}     
 Since close to threshold only a small number partial waves contribute
 to the $pp\to pp\eta$ transition amplitude, it is feasible to expand them
 in terms of angular momentum. The transition amplitude is an operator in
 spin space that depends upon the three c.m.  momenta which determine the kinematics
 of the reaction, namely the initial proton momentum $\bm{p}$, the relative
 momentum of the final state protons $\bm{k}$, and, finally the momentum
 of the eta relative to the pp pair $\bm{q}$ (other possible choices will be
 discussed later on).
 The quantum numbers associated with the
 initial state are: the angular momentum $L$, the total spin $s_i$ and the
 total angular momentum J. In the final state, the angular momentum and the 
 the total spin of the pp pair are denoted as $\ell$ and $s_f$, respectively,
 and the relative motion of the eta is described by the angular momentum $\lambda$.
 Among the above quantum numbers only J is conserved while the remaining quantum
 numbers are constrained by parity conservation and by the Pauli principle. 
 Since the eta is a pseudoscalar meson, parity conservation requires that
 $L+\ell+\lambda$ must be an odd number.
 A list of possible transitions consistent with the above restrictions 
 involving the lowest angular momenta is presented in Table \ref{table_states}.
\begin{table}
\caption{
\label{table_states}
List of allowed transitions $\mbox{}^{2s_i+1}L_J\to\mbox{}^{2s_f+1}\ell_J,\lambda$
for the lowest partial waves in the
reaction $pp\to pp\eta$. 
}
\begin{ruledtabular}
	\begin{tabular}{cc}
 even $\ell$ & odd $\ell$ \\
\hline
$\mbox{}^3 P_0\to\mbox{}^1 S_0$,s  &$\mbox{}^1 S_0\to\mbox{}^3 P_0$,s\\  
$\mbox{}^3 P_2\to\mbox{}^1 S_0$,d  &$\mbox{}^1 D_2\to\mbox{}^3 P_2$,s\\  
$\mbox{}^3 F_2\to\mbox{}^1 S_0$,d  &$\mbox{}^3 P_0\to\mbox{}^3 P_1$,p\\  
$\mbox{}^3 P_2\to\mbox{}^1 D_2$,s  &$\mbox{}^3 P_2\to\mbox{}^3 P_2$,p\\  
$\mbox{}^3 F_2\to\mbox{}^1 D_2$,s  &$\mbox{}^3 F_2\to\mbox{}^3 P_1$,p\\  
                                   &$\mbox{}^3 F_2\to\mbox{}^3 P_2$,p\\  
                                   &$\mbox{}^3 P_1\to\mbox{}^3 P_0$,p\\  
                                   &$\mbox{}^3 P_1\to\mbox{}^3 P_1$,p\\  
                                   &$\mbox{}^3 P_1\to\mbox{}^3 P_2$,p\\  
                                   &$\mbox{}^3 F_3\to\mbox{}^3 P_2$,p\\  

\end{tabular}
\end{ruledtabular}
\end{table}
 These transitions can be classified as even or odd according to the value
 of the angular momentum $\ell$ of the final state protons. In the transition matrix
 element the even (odd) partial wave amplitude will be 
 multiplied by an appropriate projection operator which
 is symmetric (antisymmetric) under the transformation $\bm{k}\to-\bm{k}$. 
 Since the final state protons are indistinguishable all observables must 
 be invariant under the interchange of the proton momenta, i.e. under the
 $\bm{k}\to-\bm{k}$ transformation. This means that 
 all interference terms between even and odd states which are antisymmetric under
 $\bm{k}\to-\bm{k}$ transformation are bound to vanish
 if one wants to respect Pauli principle. In consequence, interference
 is allowed only between partial waves belonging to the same group
 which significantly reduces the number of terms in the cross section.
 \par
 For small excitation energy the final state pp interaction appears to be
 the dominant effect and therefore it seems justified to neglect  in the cross section
 all terms which do not contain the enhancement factor C(k). With this
 assumption the only contribution to the cross section comes from the
 even partial waves listed in Table \ref{table_states}.  
 The partial wave amplitudes are functions of
 both, $k^2$ and $q^2$ but since these two quantities are linked
 by energy conservation one of them is redundant. Terms linear in k or
 q will be absent which can be seen as follows. 
 The momentum dependence in a partial wave amplitude originates from the
 spherical Bessel functions $j_\ell(k\xi)\,j_\lambda(q\eta)$ occurring
 in the appropriate overlap integrals where $\xi$ and $\eta$ are the
 Jacobi coordinates canonically conjugated with k and q, respectively. 
 For even orbital momentum
 the spherical Bessel function is an even function of the argument
 so that in the expansion of the partial wave amplitudes only even powers
 of the momenta will be admitted. However, for practical reasons,
 it does not seem feasible to go beyond the second order
 in k and q. 
 \par
 The above considerations lead us to take the following simple expression
 for the $pp\to pp\eta$ reaction cross section
 \begin{equation}
 \begin{split}
  \D\sigma /\D Lips&=|C(k)|^2\;[a+b\,P_2(\hat{\bm{p}} \cdot \hat{\bm{q}})]+\\
   +&C_0(\eta)\;[d\;\text{Re}\,C(k)+e\;\text{Im}\,C(k)]\,P_2(\hat{\bm{p}} \cdot \hat{\bm{k}}).\\
	 \label{t1}
 \end{split}
 \end{equation}
 In \eqref{t1} we are using the standard notation where
  $\D Lips$ stands for the invariant three-body phase space element.
  In \eqref{t1} a denotes the modulus squared of the sole production amplitude which 
 survives at threshold, associated with the transition $\mbox{}^3 P_0\to\mbox{}^1 S_0,s$,
  b represents the interference term between the latter amplitude  
  and the $\mbox{}^3 P_0\to\mbox{}^2 S_0,d$ amplitude, and, the (d,e) pair describes,
  respectively, the interference with the $\mbox{}^3 P_2\to\mbox{}^1 D_2,s$ amplitude.
  All of the above mentioned coefficients are functions of the final state momenta.
  Taking into account only threshold behavior, we can see that a is constant,
  b will be proportional
  to $q^2$ and (d,e) both are of the order of $k^2$. In this situation, we have to
  expand  a to the same order setting $a=a_0+a_1\,q^2$ where $a_0$ and $a_1$ are
  two unknown parameters. The parameter $a_0$ can be absorbed in the normalization
  constant adjusted to the experimental total cross section value, and the 
  ultimate expression for the cross section reads
 \begin{equation}
 \begin{split}
	 \frac{\D\sigma}{\D Lips}& \propto |C(k)|^2\;
	\left \{1+\dfrac{q^2}{q_{max}^2}\,
	\left [x+y\,P_2(\hat{\bm{p}}\cdot\hat{\bm{q}})\right] \right\}+\\
	 +\dfrac{k^2}{k_{max}^2} & C_0(\eta)\;\left[z_r\;\text{Re}\,
    C(k) + z_i\;\text{Im}\,C(k)\right]\,P_2(\hat{\bm{p}}\cdot\hat{\bm{k}}),\\
	 \label{t2}
 \end{split}
 \end{equation}
 where $x,y,z_r,z_i$ are real dimensionless parameters to be determined. For reasons
 of convenience we have scaled q and k by dividing them by their maximal values
 $q_{max}$ and $k_{max}$, respectively. The parameter x  represents the correction
 of the order $q^2$ to the dominant transition, the parameters y and $(z_r, z_i)$
 provide the measure of the admixture of the d-waves.

 \section{\label{sec:3}Comparison with experiment}
\begin{figure}
\begin{center}
\includegraphics[scale=0.4]{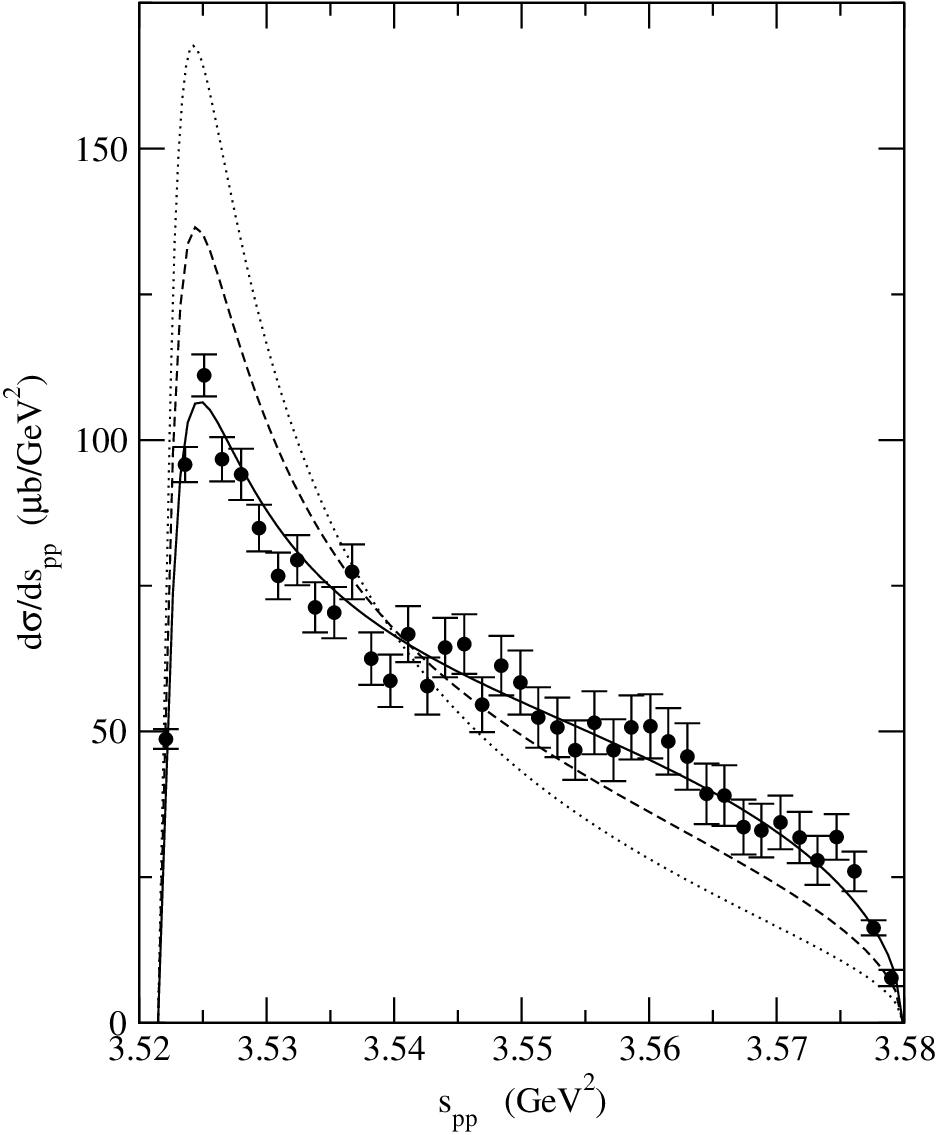}
\end{center}
\caption{$s_{pp}$ distribution is compared with the data form Moskal et al. \cite{moskal}.
The dotted curve corresponds to a constant 
$\mbox{}^3 P_0\to\mbox{}^1 S_0\, s$ transition amplitude with on-shell enhancement factor
\eqref{t10a} calculated from  \eqref{t11}. 
In the dashed curve the complete  $|C(k)|^2$  has been used
while in the full curve, at the top of that,
a correction linear in energy has been included.   }
\label{fig:3}
\end{figure}

\begin{figure}
\begin{center}
\includegraphics[scale=0.4]{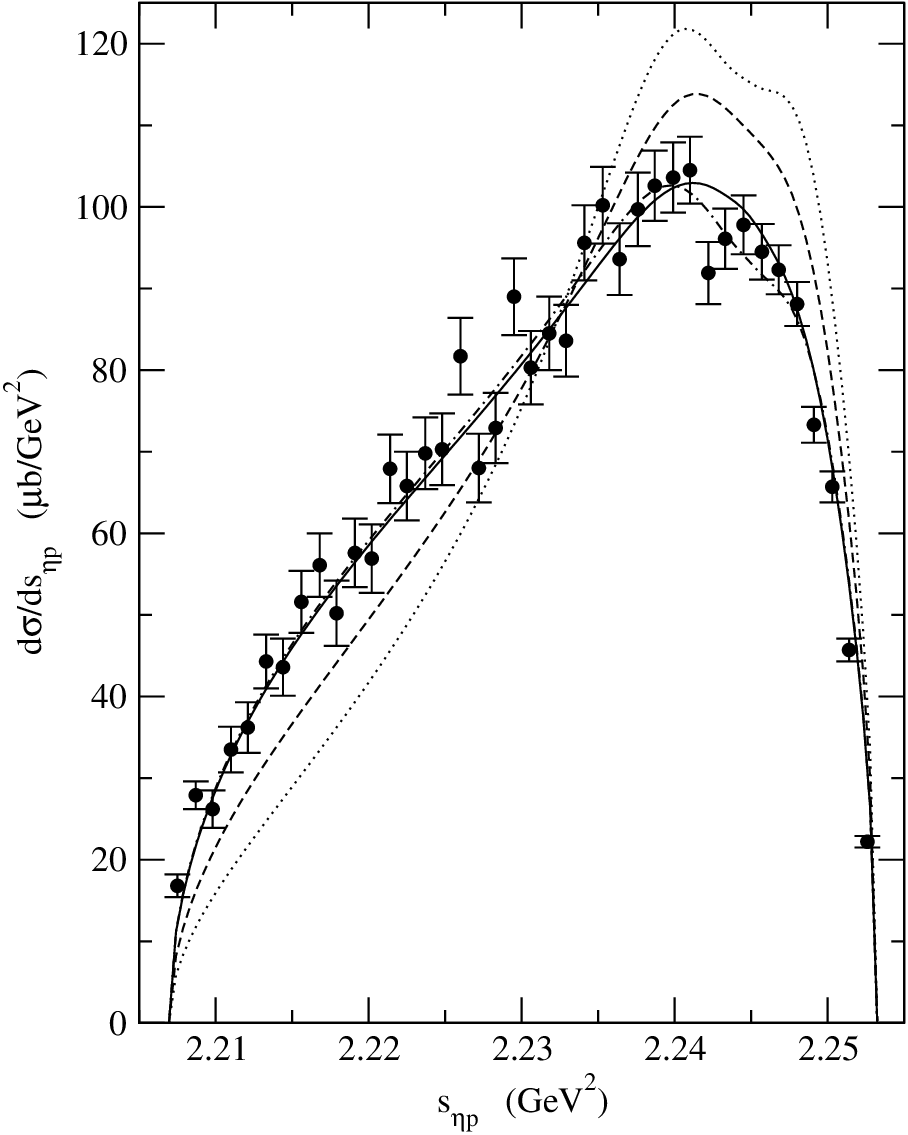}
\end{center}
\caption{$s_{\eta p}$ distribution is compared with the data
from Moskal et al. \cite{moskal}. 
The dotted curve  has been obtained 
keeping only 
the dominant $\mbox{}^3 P_0\to\mbox{}^1 S_0\, s$ transition amplitude
which is assumed to be a constant with  $|C_{WM}(k)|^2$ calculated from \eqref{t11}.
In the dashed curve the complete  $|C(k)|^2$  has been used
while in the dot-dashed curve, at the top of that,
a correction linear in energy has been included.   
The full curve 
accounts for also the interfernce with 
$\mbox{}^3 P_2\to\mbox{}^1 S_0\, d$ amplitude.}
\label{fig:4}
\end{figure}

 In Fig. \ref{fig:3} and in Fig. \ref{fig:4} 
 we present the experimental distribution  \cite{moskal} of
 the square of  the effective pp mass $s_{pp}$ and the 
 square of the $\eta$p mass $s_{\eta p}$, respectively.
 For a start, we
 assumed a constant production amplitude 
 setting  $x=y=z_r=z_i=0$.
 Using the on-shell enhancement factor $|C_{WM}(k)|^2$ specified in \eqref{t10a} with the 
 effective range approximation \eqref{t11}, from \eqref{t2}
 we obtain the theoretical distributions depicted by a dotted
 curve in Figs. \ref{fig:3} and \ref{fig:4}. 
 Qualitatively, the dotted curves in Figs.~\ref{fig:3} and \ref{fig:4}  
 are roughly in accord with experiment
 but they are far not satisfactory in quantitative terms.
 In both figures the peak in the calculated curves is too big
 and partially responsible for that is the oversimplified enhancement factor. 
 As we already know,   $|C_{WM}(k)|^2$ calculated by using effective range formula \eqref{t11}
 exhibits a too steep a fall and
 by normalizing the distributions to the total cross-section,
 to compensate that, the curves are moved up so that the peak gets magnified. 
 Therefore, the simplest remedy is to abandon the on-shell approximation 
 and from now on we will
 use the complete enhancement factor calculated from \eqref{t10}. 
 The appropriate distributions are presented by the dashed curves
 in Figs.~\ref{fig:3} and \ref{fig:4}.
 Although the resulting corrections go in the right direction
 bringing the calculation  closer to  experiment, but
 this is still  not enough for providing a full understanding of the data.
 In this situation, it is interesting to know
 how much we can improve the agreement
 by disposing of the various corrections discussed in the preceding section
 and represented by the adjustable parameters $x,\,y,\,z_r,\,z_i$. 
 Before proceeding, however,  it should be observed  that
 in the case of the pp effective mass distribution a major simplification
 takes place because, as  apparent from \eqref{t2},
 after  integrating over the angles only the
 s-wave amplitude survives. Therefore, the parameters $y,\,z_r,\,z_i$ 
 representing interference terms never enter $s_{pp}$ distribution. 
 \printfigures
 By contrast,  in the case of $s_{\eta p}$ distribution
 the interference terms in general do not vanish in result of angular
 integration. Since a non-relativistic approach is here
 completely adequate, in order to obtain
 the cross section as a function of $s_{\eta p}$ we have to introduce 
 an equivalent Jacobi frame taking as the two independent momenta 
 the $\eta$-p relative momentum $\bm{k}_1$ and the momentum of the other proton
 relative to this pair $\bm{q}_1$. The transformation between the two Jacobi
 frames,  takes the form
 \begin{equation}
 \binom{\bm{k}}{\bm{q}}=
           \begin{pmatrix}
                   -1/2  & -\mu/\nu\\      
                     1   & -\mu/M         
           \end{pmatrix}
    \binom{\bm{k}_1}{\bm{q}_1}
	 \label{t3}
 \end{equation}
 where $\mu$ is the eta-p reduced mass and $\nu$ is the reduced mass of the eta and
 the pp pair. Clearly, substituting \eqref{t3} in \eqref{t2} and integrating over
 the angles of $\bm{k}_1$ and $\bm{q}_1$, the interference terms give non-vanishing 
 contribution. 
 \par
 Since the $s_{pp}$ distribution depends solely upon  x, we
 adjusted this parameter by fitting the theoretical $s_{pp}$ 
 distribution to the experimental data with the best fit value of x being  x=-0.514.
 The resulting distribution  displayed in Fig.\ref{fig:3} by the full curve
 agrees now quite well with experiment. Actually, since the interference terms drop out
 and the $\eta$p interaction gives here a small effect, there is really not much room
 for improvement other than  including the $q^2$ correction in the  
 the dominant $\mbox{}^3 P_0\to\mbox{}^1 S_0\, s$ transition amplitude.
 Next, with the value of x in hand,
 we  calculated the $s_{\eta p}$ distribution, still leaving out the interference terms
 i.e., setting  $y=z_r=z_i=0$.
 The resulting cross section which does not involve adjustable parameters
 any more is presented in Fig.\ref{fig:4} by a dot-dashed curve. We can see that
 also in this case the agreement with experiment looks much better. 
 The interference terms in \eqref{t2} involving the parameters $(z_r,\,z_i)$  are probably
 much smaller than the term proportional to y because the former terms exhibit
 only a linear dependence upon the enhancement factor, so we ignored these terms 
 adjusting the single parameter y to the experimental $s_{\eta p}$ distribution.
 With  x fixed, the best fit value of y was y=3.38. The corresponding cross section
 is displayed in Fig. \ref{fig:4} by the full curve and the agreement is quite
 good. We tried also to vary the parameters $(z_r,\,z_i)$ but this was not 
 very successful as the fitting procedure attempts to reproduce a structure
 at the high energy end of the spectrum. With  three adjustable parameters
 it is quite easy to produce a two-peaked distribution with a good $\chi^2$.
 Since it is not  quite certain that the data really reveal a two-peaked distribution
 we did not pursue this fit any further. It should be also mentioned that when
 the interference term linear in C(k) is accounted for, the calculation becomes very
 sensitive to the NN potential used to obtain C(k) because the normalization
 of C(k) can not be absorbed in the overall normalization of the cross-section.
 \par
\begin{figure}
\begin{center}
\includegraphics[scale=0.4]{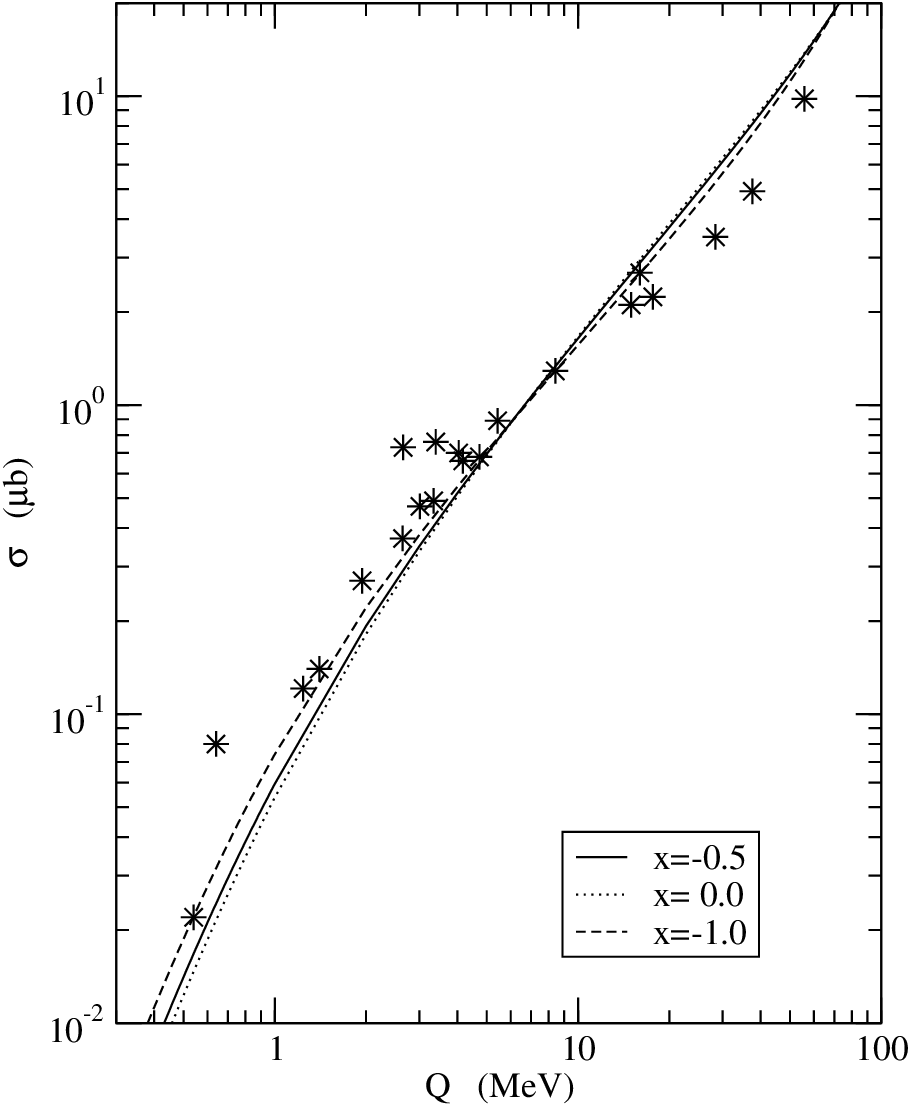}
\end{center}
\caption{The total cross section for the $pp\to pp\eta$ reaction versus the
excitation energy Q for different values of the parameter x.
The experimental data are from \cite{exp}.   }
\label{fig:5}
\end{figure}

 In Fig.~\ref{fig:5} we are going to present the changes in the excitation
 energy distribution caused by adding an energy 
 dependent term to the eta production amplitude 
 (the interference terms in \eqref{t2} give no contribution). 
 \printfigures
 The case x=0 corresponds to a constant production amplitude (dotted curve in
 Fig.~\ref{fig:5}), while x=-1 (dashed curve in Fig.~\ref{fig:5}) 
 represents the lower limit for this parameter which must be imposed  
 to prevent the cross section from going negative. The best fit value x$\approx$-0.5
 necessary to reproduce the effective mass distributions for Q=15.5~MeV is roughly
 midway between these values and the corresponding cross section is presented
 by the full curve in Fig.~\ref{fig:5}. The linear energy dependence introduced
 in the amplitude of the eta production has little impact on  
 the total cross section. In fact, the two curves where x$\le$0 exhibit slightly better agreement
 with experiment as compared with the dotted curve corresponding to constant production amplitude.
 To obtain our curves we used the full expression for the enhancement factor.
 We wish to note that as a result of 
 using the on-shell approximation for the enhancement factor
 the calculated cross section is underestimated at large Q. 
 \par
 With a good fit to the effective mass distributions for Q=15.5~MeV,
 it is interesting to know whether   
 the approach presented above makes sense for Q=41~MeV since for this value of the excitation
 energy the experimental data are also  available  in the literature \cite{zupran}.
 Strictly speaking, for this much larger excitation energy the validity of the simple 
 formula \eqref{t2} becomes questionable and the inclusion of p-waves might be indispensable.
 Nevertheless, it is useful to find out what a simple model can do for larger Q. 
\begin{figure}
\begin{center}
\includegraphics[scale=0.4]{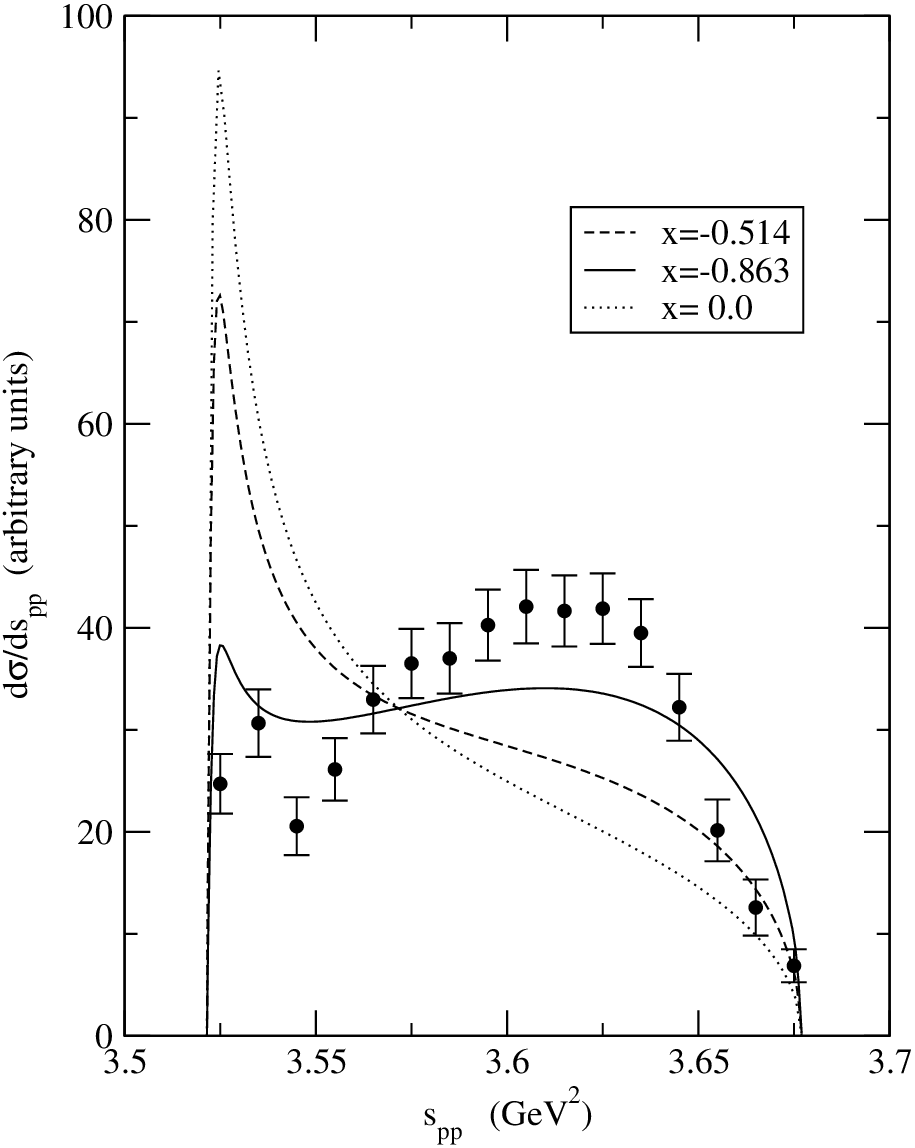}
\end{center}
\caption{$s_{pp}$ distribution for an excitation energy Q=41~MeV 
is compared with the data form Abdel-Bary et al. \cite{zupran}.
In our calculation  we have neglected all higher waves except 
the $\mbox{}^3 P_0\to\mbox{}^1 S_0\, s$ transition amplitude. 
 The energy dependence in the latter is proportional to the parameter x,
 cf. \eqref{t2}.}
\label{fig:6}
\end{figure}
 \printfigures
\begin{figure}
\begin{center}
\includegraphics[scale=0.4]{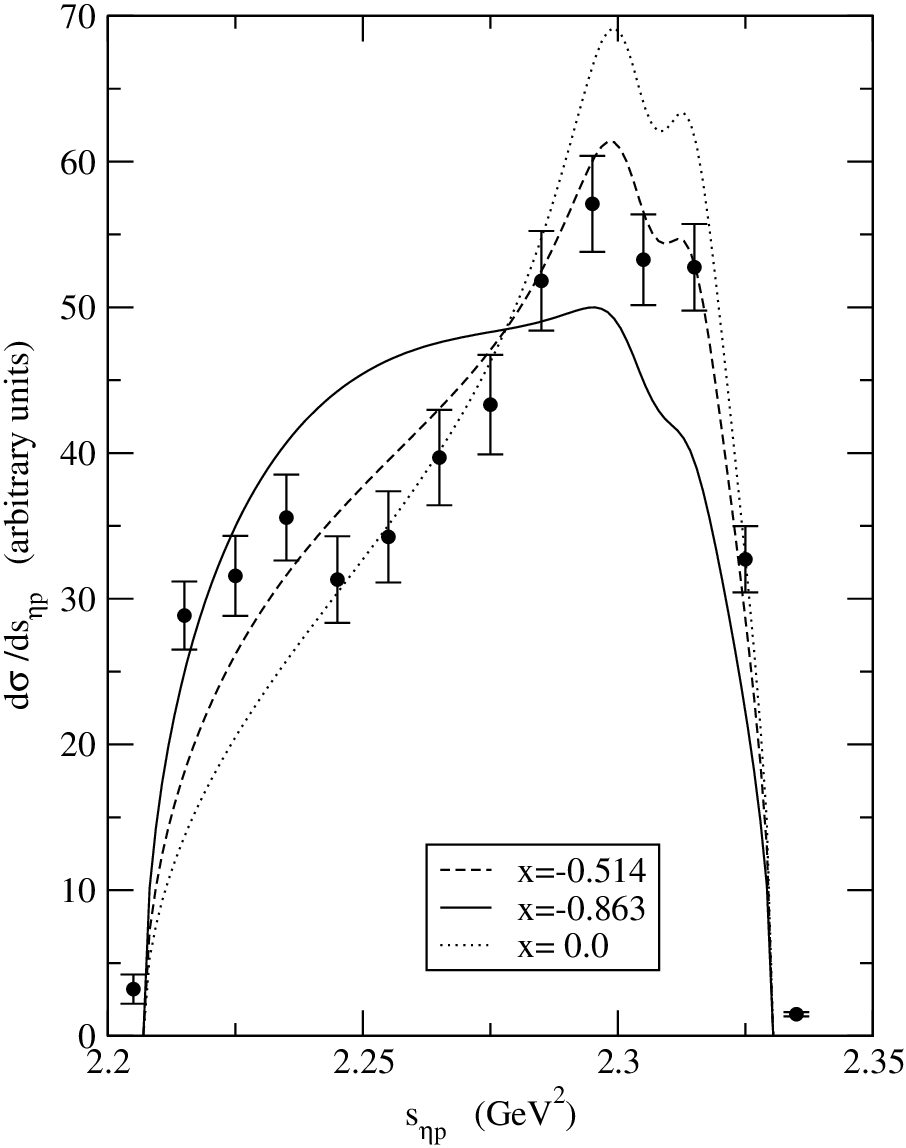}
\end{center}
\caption{$s_{\eta p}$ distribution for an excitation energy Q=41~MeV 
is compared with the data form Abdel-Bary et al. \cite{zupran}.
In our calculation  we have neglected all higher waves except 
the $\mbox{}^3 P_0\to\mbox{}^1 S_0\, s$ transition amplitude. 
 The energy dependence in the latter is proportional to the parameter x,
 cf. \eqref{t2}.}
\label{fig:7}
\end{figure}
 \printfigures
 The results of our computations are presented in Figs.~\ref{fig:6} and \ref{fig:7}
 where they are compared with the data from \cite{zupran}. In the calculation
 we neglected the interference terms, setting $y=z_r=z_i=0$ in \eqref{t2}.
 The dotted curve in Fig.~\ref{fig:6} and Fig.~\ref{fig:7} corresponds to a
 constant eta production amplitude. Whilst  the  
  $s_{\eta p}$ distribution is not too far off the experiment,   the 
  $s_{pp}$ distribution is in a bad shape as the peak caused by pp FSI is definitely
  much too big. The dashed curve is obtained by adopting for x the same value that
  for Q=15.5~MeV gave the best agreement with experiment i.e., we are pretending that this
  parameter does not vary with Q. This brings improvement in both cross sections
  and in fact the $s_{\eta p}$ distribution agrees with the data quite well.
  Since the parameter x may depend upon Q, and for Q=41~MeV its value could be different
  than for Q=15.5~MeV,
  we allowed x to vary, adjusting its value to the data from  \cite{zupran}.
  This time the best fit value turns out to be x=-0.863 and the corresponding distributions
   are presented by  full curves in Figs.~\ref{fig:6} and \ref{fig:7}.
  It is not possible to improve the agreement in
  both distributions using the same value of the parameter x, and
  therefore the fit depicted by the full curve is a compromise. Although the corrections
  go in the right direction but the agreement is still not satisfactory and we have
  to accept the fact that
  for Q=41~MeV the assumed functional form of the cross-section is incomplete.
  We tried to include the interference terms with the d-waves but the fit was
  unsuccessful. Clearly, further extensions call for additional parameters but this
  does not seem to be affordable with the present data, and therefore we stop here.  
 \par
  Summarizing,  a simple model in which the $pp\to pp\eta$
  cross section contains only the dominant  
  $\mbox{}^3 P_0\to\mbox{}^1 S_0,s$ partial wave amplitude
  corrected by the final state pp interaction is capable of explaining
  the current experimental data at an excitation energy Q=15.5~MeV.
  The experimental pp and the $\eta$p effective mass distributions can be both
  reproduced  if: 
  (i) in the eta production amplitude 
    a correction linear in energy  is admitted 
   and, (ii)  
   the full pp FSI enhancement factor without the on-shell approximation is used.
   In the presented calculation  the $\eta$p final state interaction has not been
   included explicitly for reasons outlined in the Introduction
   but since the parameters x and y have been adjusted to the data, they
   effectively account for this effect. 
  
 \begin{acknowledgments}
  The author wishes to thank S.~Wycech for reading the manuscript  and 
  for constructive criticism.
  Partial support under grant KBN 5B 03B04521 is gratefully acknowledged. 
 \end{acknowledgments}

 \end{document}